\documentclass[twocolumn,preprintnumbers,amsmath,amssymb]{revtex4-1}

\usepackage[version=3]{mhchem} 
\usepackage{amssymb}
\usepackage{graphicx}
\usepackage{dcolumn}
\usepackage{braket}
\usepackage{gensymb}
\usepackage{color}
\usepackage{footnote} 

\begin{document}

\title{Dynamics of solvation and desolvation of rubidium attached to He nanodroplets}

\author{J. von Vangerow}
\thanks{O. John and J. von Vangerow contributed equally to this work}
\author{O. John}
\thanks{O. John and J. von Vangerow contributed equally to this work}
\author{F. Stienkemeier}
\author{M. Mudrich}


\affiliation{Physikalisches Institut, Universit{\"a}t Freiburg, 79104 Freiburg, Germany}

\date{\today}

\begin{abstract}
The real-time dynamics of photoexcited and photoionized rubidium (Rb) atoms attached to helium (He) nanodroplets is studied by femtosecond pump-probe mass spectrometry. While excited Rb atoms in the perturbed 6p-state (Rb$^\ast$) desorb off the He droplets, Rb$^+$ photoions tend to sink into the droplet interior when created near the droplet surface. The transition from Rb$^+$ solvation to full Rb$^\ast$ desorption is found to occur at a delay time $\tau\sim 600$ fs for Rb$^\ast$ in the 6p$\Sigma$-state and $\tau\sim 1200$ fs for the 6p$\Pi$-state. Rb$^+$He ions are found to be created by directly exciting bound Rb$^\ast$He exciplex states as well as by populating bound Rb$^+$He-states in an photoassociative ionization process.

\end{abstract}

\maketitle

\section{\label{sec:Intro}Introduction}
The laser-induced dynamics of pure and doped helium (He) nanodroplets is currently attracting considerable attention~\cite{Mudrich:2008,Gruner:2011,KrishnanPRL:2011,Kornilov:2011,PentlehnerPRL:2013,Ovcharenko:2014,Mudrich:2014}. While the superfluidity of He nanodroplets has been tested in numerous key experiments by probing stationary properties~\cite{Hartmann2:1996,Grebenev:1998,Brauer:2013}, the impact of the quantum nature of the droplets on their dynamic response to impulsive excitation or ionization is much less well established. As a prominent recent example, the rotational dynamics of various molecules embedded in He droplets induced by impulsive alignment was found to be significantly slowed down and rotational recurrences were completely absent~\cite{PentlehnerPRL:2013}. This indicates that substantial transient system-bath interactions are present during the laser pulse. In contrast, the vibrational dynamics of rubidium (Rb) molecules Rb$_2$ attached to the surface of He nanodroplets revealed only slow relaxation and dephasing proceeding on a nanosecond time scale~\cite{Mudrich:2009,Gruner:2011}.

Various recent experimental and theoretical studies have addressed the dynamics of solvation and desolvation of ionized or excited metal atoms off the surface of He nanodroplets~\cite{Loginov:2007,Loginov:2012,Fechner:2012,Zhang:2012,Vangerow:2014,Loginov:2014,Mudrich:2014,TheisenImmersion:2011,Theisen:2010,Mateo:2013,Leal:2014}. So far, these studies have concentrated on measuring the total yield and the final velocity of the ejected atoms as a function of the atomic species and the electronic state of excitation. In this paper we present the first time-resolved characterization of the desorption process of Rb atoms off the surface of He nanodroplets upon excitation to the droplet-perturbed states correlating to the 6p atomic orbital. The experimental scheme we apply is femtosecond (fs) pump-probe photoionization in combination with time-of-flight mass-spectrometry. We find that the yield of detected Rb$^+$ photoions as a function of delay time $\tau$ between the exciting pump and the ionizing probe pulses is determined by the interplay of the repulsive interaction of excited Rb$^\ast$ with respect to the He surface and the attractive interaction of the Rb$^+$ ion with the He surface induced by photoionization.

The Rb$^\ast$-He droplet repulsion initiates the desorption of the Rb$^\ast$ atom off the He droplet surface. Except for the lowest excited state of Rb, 5p$_{1/2}$, all excited states up to high Rydberg levels experience strong repulsion from He droplets~\cite{Aubock:2008,Callegari:2011}. In contrast, the Rb$^+$-He droplet attraction causes the Rb$^+$ ion to fall back into the He droplet when created near the He droplet surface at short delay times~\cite{Theisen:2010,Leal:2014}. Atomic cations are known to form stable ``snowball'' structures consisting of a cationic core which is surrounded by a high density shell of He atoms. As a result, free Rb$^+$ ions appear in the mass spectrum only after a characteristic pump-probe delay time $\tau_D$, which depends on the state the Rb atom is initially excited to. 


In addition to neat Rb$^+$ atomic ions, the photoionization mass spectra contain Rb$^+$He molecular ions in the full range of laser wavelengths correlating to the droplet-perturbed Rb 6p-state. The occurrence of such molecular ions has previously been interpreted by the formation of metastable `exciplex' molecules~\cite{Droppelmann:2004,Mudrich:2008,Giese:2012,Fechner:2012,Vangerow:2014,Mudrich:2014}. These bound states of excited metal atoms and one or few He atoms can be populated either by a tunneling process~\cite{Reho:2000,Reho2:2000,Loginov:2007,Loginov:2015} or by direct laser-excitation of bound states in the metal atom-He pair potential~\cite{Pascale:1983,Fechner:2012,Vangerow:2014,Loginov:2014}. In the former case, exciplex formation times $\gtrsim 50$~ps are expected~\cite{Reho2:2000,Droppelmann:2004}, whereas in the latter case, exciplexes are created instantaneously. Thus, previous pump-probe measurements revealing exciplex formation times of $8.5$ and $11.6$~ps for Rb$^4$He and Rb$^3$He, respectively, upon excitation into the droplet-perturbed 5p$_{3/2}$-state could not be consistently interpreted~\cite{Droppelmann:2004}. 

In the present study we observe a time-delayed increase of the Rb$^+$He signal as for Rb$^+$ indicating that the pump-probe dynamics is primarily determined by the competition between desorption of the Rb$^\ast$He exciplex off the He droplet surface and the Rb$^+$He cation falling back into the He droplet interior. 
Moreover, a pronounced maximum in the Rb$^+$He signal transients indicates that an additional Rb$^+$He formation channel besides photoionization of Rb$^+$He exciplexes is active -- photoassociative ionization (PAI) of the desorbing Rb atom and a He atom out of the droplet surface. PAI is a well-known process where a bound cationic molecule or complex is formed by photoionization or photoexcitation into autoionizing states of an atom or molecule of a collision complex~\cite{Shaffer:1999}. PAI is a special case of traditional associative ionization where a bound molecular cation is formed in a binary collision of an electronically excited atom~\cite{Weiner:1990}. In either case the binding energy is taken away by the electron emitted in the process.

\section{Experimental setup}
The experimental setup is similar to the previously used arrangement~\cite{Mudrich:2009,Fechner:2012} except for the ionization and detection schemes. He droplets are produced by a continuous supersonic expansion of He~6.0 through a 5~$\mu$m nozzle at a pressure of $50$~bar. The transversal velocity spread of the beam is reduced by placing a 400~$\mu$m skimmer 13~mm behind the nozzle. Unless otherwise stated, the nozzle temperature is kept at 17~K. This results in a log-normal distribution of the He droplet size with a mean size of $1.1\times 10^4$ He atoms. Subsequently, the droplet beam passes a mechanical chopper and a Rb-filled cell of length 1~cm, stabilized at a temperature of $85~\degree$C. At the corresponding vapor pressure, most droplets pick up on average one Rb atom following poissonian statistics. By overlapping the droplet beam with the output of the fs laser, we resonantly excite and ionize the dopant atom. 

In contrast to previous studies, we use amplified fs laser pulses generated by a regenerative amplifier operated at a pulse repetition rate of 5 kHz. At this repetition rate, multiple excitations of Rb atoms by subsequent pulses from the pulse train are safely excluded. The pulses are frequency-doubled in a BBO crystal resulting in a pulse duration of $t_p=120$~fs with a variation for different laser center wavelengths of 20~fs. Two identical, time-delayed pump and probe pulses are generated by means of a mechanical delay line. The laser beam is focused into the vacuum chamber using a 30~cm lens which leads to a peak intensity in the range of $5\times 10^{12}$ Wcm$^{-2}$. 

Photoions are detected by a time-of-flight (TOF) mass spectrometer in Wiley-McLaren configuration mounted in-line with the He droplet beam~\cite{Wiley:1955}. At the end of the drift tube a high negative potential is applied to further accelerate the arriving ions which boosts the efficiency of detecting large cluster masses in the $10^4$~amu range using a Daly-type detector~\cite{Daly:1960}. The latter consists of a Faraday cup, a scintillator with an optical bandpass interference filter and a photomultiplier tube. In case of electron detection, a simple electrode setup consisting of a repeller, an extractor grid and a channeltron detector with positive entrance potential is used. For both detectors, the resulting pulses are amplified, threshold-discriminated and acquired by a fast digitizer. When detecting heavy masses a counting unit is used.             

\section{R\MakeLowercase{b} desorption dynamics}
In the present paper we concentrate on the fs pump-probe dynamics of Rb atoms attached to He nanodroplets which are excited to droplet-perturbed states correlating to the atomic 6p-state. These states have previously been studied using nanosecond pulsed excitation and velocity-map imaging of photoions and electrons~\cite{Fechner:2012,Vangerow:2014}. Due to the interaction of the excited Rb atom with the He droplet surface, the 6p-state splits up into the two states 6p$\Sigma$ and 6p$\Pi$ according to the pseudo-diatomic model which treats the whole He droplet, He$_N$, as one constituent atom of the RbHe$_N$ complex~\cite{Stienkemeier:1996,LoginovPRL:2011,Callegari:2011}. 

\begin{figure}
\centering
\includegraphics[width=0.45\textwidth]{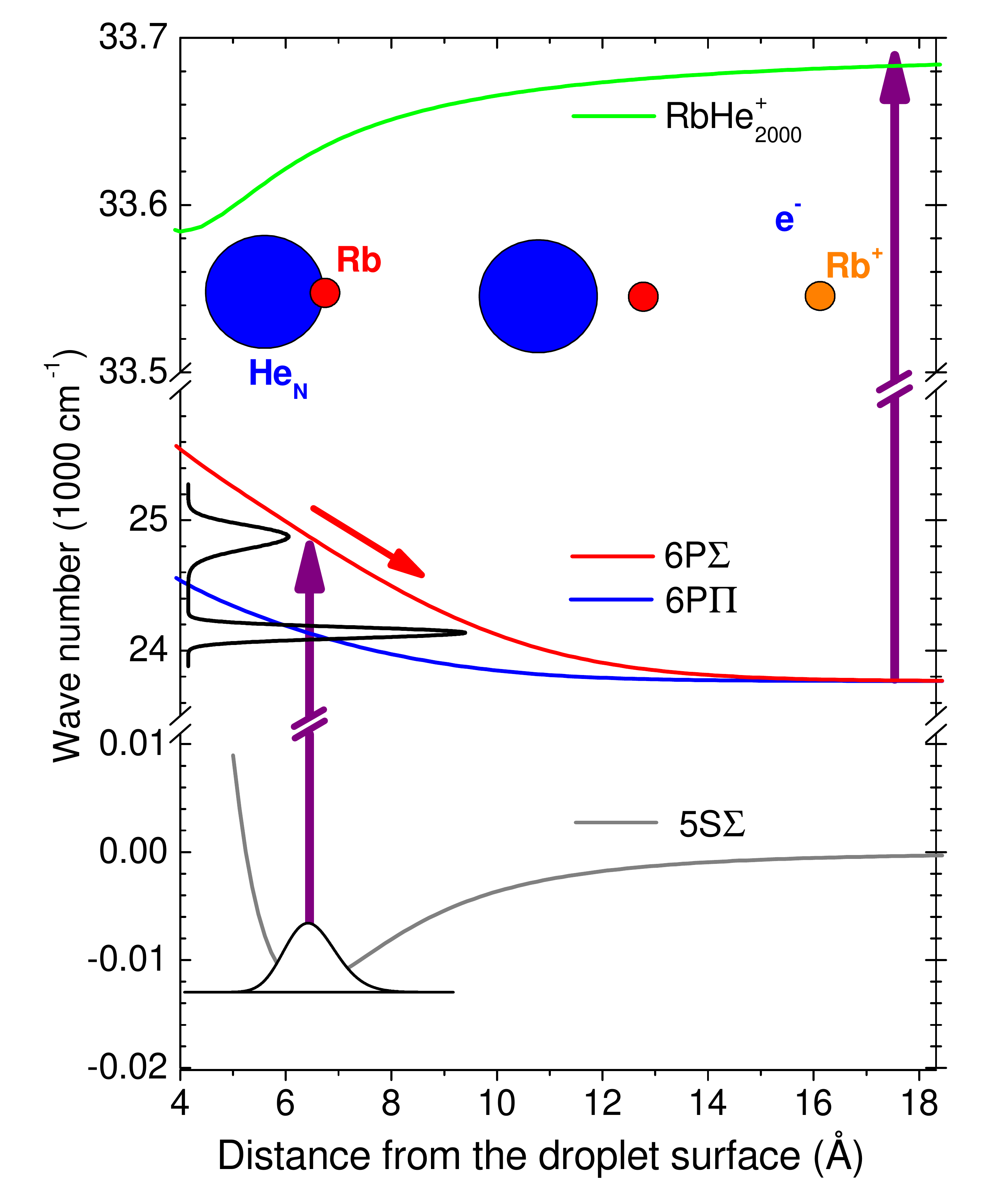}
\caption{Potential energy diagram of the Rb-He nanodroplet complex. Vertical arrows depict the photo-excitation and ionization processes. The potential curves of the neutral Rb-He$_{2000}$ complex are taken from~\cite{Callegari:2011}, the one of the Rb$^+$-He$_{2000}$ complex is obtained from the Rb$^+$-He pair potential~\cite{Koutselos:1990} on the basis of the He density distribution of the groundstate RbHe$_{2000}$ complex~\cite{Pi}. The two peaks plotted vertically on the left-hand scale show the expected excitation spectrum based on these potentials.}
\label{fig:potentials}
\end{figure}

Using the RbHe$_N$ pseudo-diatomic potential curves for the 5s$\Sigma$ electronic groundstate and the 6p$\Sigma,\,\Pi$ excited states we compute the Franck-Condon profiles for the expected vertical excitation probability using R. LeRoy's program BCONT~\cite{bcont}. The corresponding transition probability profile is depicted on the left-hand side of Fig.~\ref{fig:potentials}. The experimental excitation spectrum is in good agreement with the calculated one apart from the fact that the experimental peaks are somewhat broader~\cite{Fechner:2012}. Since both 6p$\Sigma$ and 6p$\Pi$ pseudo-diatomic potentials are shifted up in energy by up to 1200 cm$^{-1}$ with respect to the atomic 6p level energy, we expect strong repulsion and therefore fast desorption of the Rb atom off the He droplet surface to occur following the excitation. 

However, upon ionization of the excited Rb atom by absorption of a second photon (vertical arrow on the right-hand side of Fig.~\ref{fig:potentials}), the interaction potential suddenly turns weakly attractive. Thus, the Rb$^+$ ion may be expected to turn around and to fall back into the He droplet provided ionization occurs at short delay times after excitation such that the desorbing Rb$^\ast$ picks up only little kinetic energy
\begin{equation}
E_{kin,\,\mathrm{Rb}^\ast}(R)<E_{pot,\,\mathrm{Rb}^+}(R).
\label{eq:ineq}
\end{equation}  
Here, $E_{pot,\,\mathrm{Rb}^+}(R)$ denotes the lowering of the potential energy of the Rb$^+$ ion due to the attractive interaction with the He droplet at the distance $R$ from the droplet surface. Eq.~\ref{eq:ineq} holds for short distances $R<R_{c}$ falling below a critical value $R_{c}$. When assuming classical motion, we can infer from Eq.~\ref{eq:ineq} the critical distance $R_{c}$ for the turn-over. From simulating the classical trajectory $R(t)$ we can then obtain the delay time $\tau_c$ at which the turn-over occurs. In the following we refer to $\tau_c$ as `fall-back time'. Thus, when measuring the number of free Rb$^+$ ions emitted from the He droplets by pump-probe photoionization we may expect vanishing count rates at short delays $\tau <\tau_c$ due to the Rb$^+$ ions falling back into the droplets, followed by a steep increase and subsequent constant level of the Rb$^+$ signal at delays $\tau>\tau_c$. 

\subsection{Experimental results}
\begin{figure}
\centering
\includegraphics[width=0.45\textwidth]{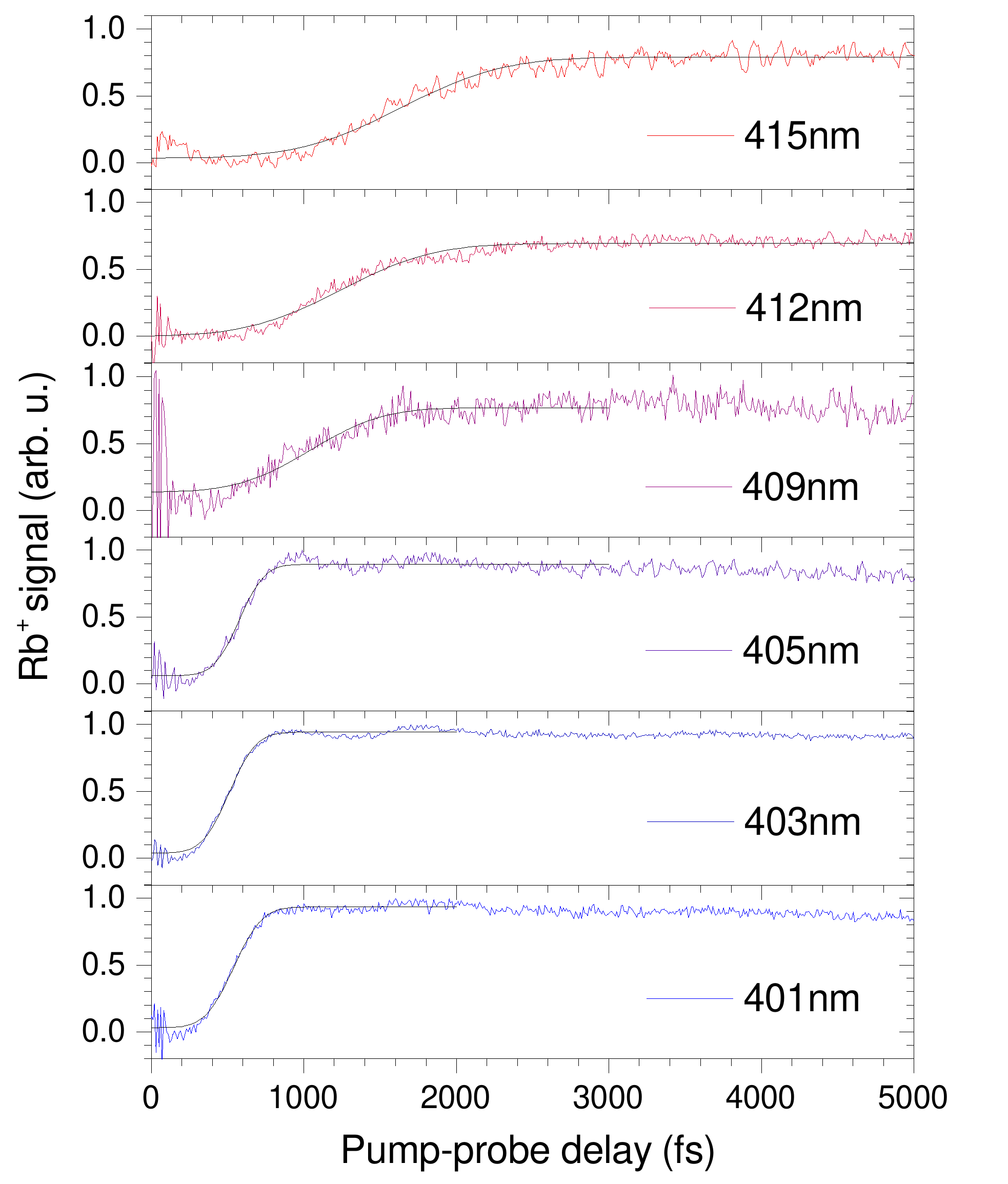}
\caption{Pump-probe transient Rb$^+$ ion count rates recorded for various wavelengths $\lambda$ of the fs laser. At $\lambda\gtrsim 409$ nm, excitation occurs predominantly to the 6p$\Pi$-state, at $\lambda\lesssim 409$ nm predominantly the 6p$\Sigma$-state is excited. The thin smooth lines are fits to the data.}
\label{fig:RbTransients}
\end{figure}
Fig.~\ref{fig:RbTransients} shows the transient Rb$^+$ ion signals measured by integrating over the $^{85}$Rb and $^{87}$Rb mass peaks in the time-of-flight mass spectra recorded for each value of the pump-probe delay. The shown data are obtained by subtracting from the measured ion signals the sum of ion counts for pump and probe laser pulses only. The error bars stem from error propagation taking into account the uncertainties associated with the different signal contributions. By tuning the wavelength of the fs laser $\lambda$ we can excite predominantly the 6p$\Pi$ ($\lambda\gtrsim 409$ nm) or the 6p$\Sigma$-states ($\lambda\lesssim 409$) of the RbHe$_N$ complex. As expected, we observe a step-like increase of the Rb$^+$-yield at delays ranging from 600 fs ($\lambda =401$ nm) up to about 1500 fs ($\lambda =415$ nm). The signal increase occurs at shorter delays when exciting into the more repulsive 6p$\Sigma$-state because the Rb atom moves away from the He droplet surface faster than when it is excited into the shallower 6p$\Pi$-state. The rising edge of the signal jump is extended over a delay  period of about 400~fs, partly due to the finite length and bandwidth of the laser pulses. Desorption along the 6p$\Pi$-potential appears as an even smoother signal rise, indicating that a purely classical model is not suitable for reproducing the observed dynamics. For laser wavelengths $\lambda<409$ nm we observe a weakly pronounced double-hump structure with maxima around 800 and 1800~fs, respectively, which we discuss in section~\ref{sec:simulations}. 

\begin{figure}
\centering
\includegraphics[width=0.48\textwidth]{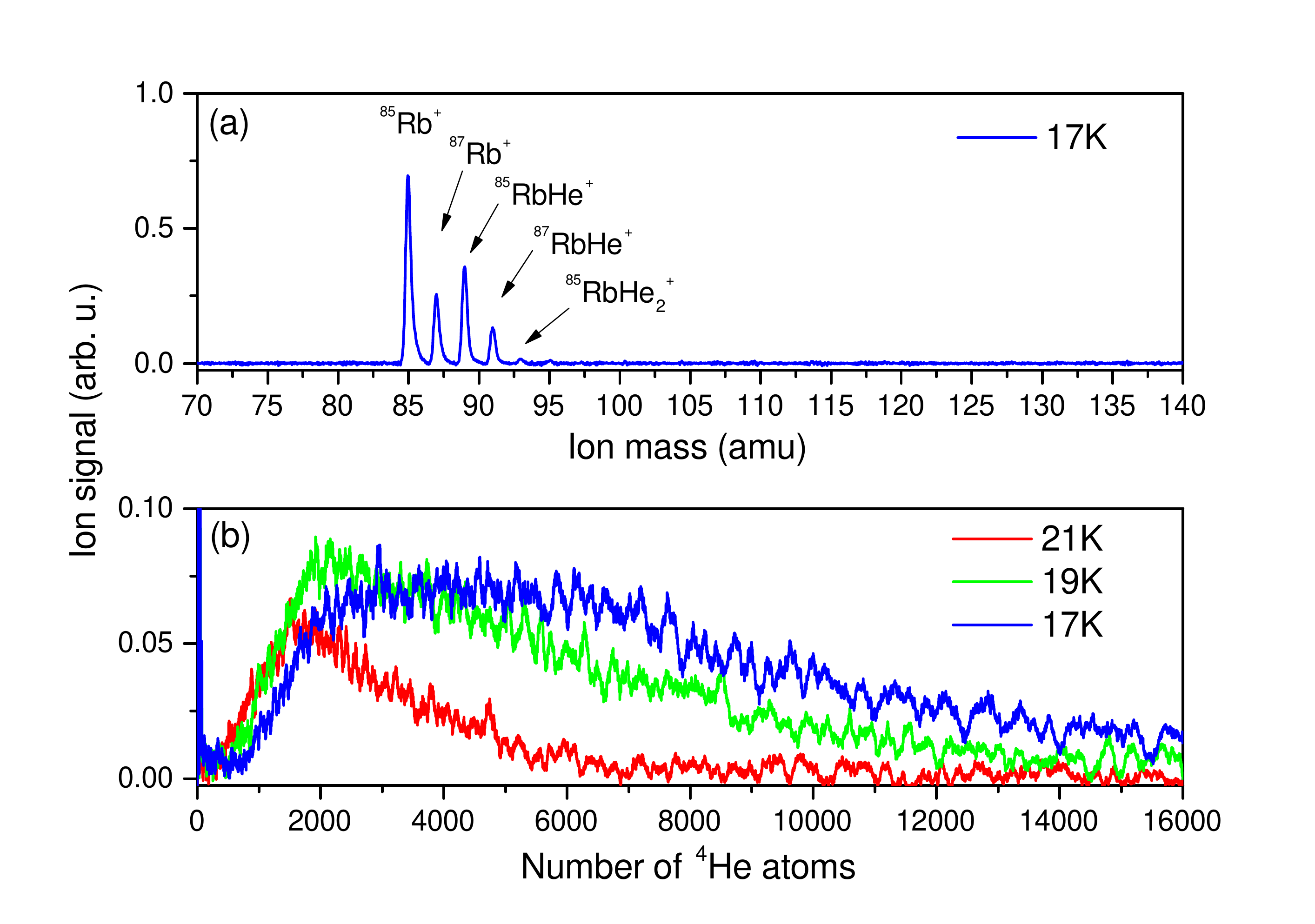}
\caption{(a) Typical mass spectra recorded for Rb-doped He nanodroplets by fs photoionization taken at a center wavelength $\lambda=415$ nm and a 5~ps pump probe delay. In addition to the atomic isotopes $^{85}$Rb$^+$ and $^{87}$Rb$^+$ the mass spectra contain Rb$^+$He and Rb$^+$He$_2$ molecular ions. (b) An extended view of mass spectra taken at various nozzle temperatures using single fs pulses at $\lambda=415$ nm reveals the presence of large masses of unfragmented ion-doped He droplets Rb$^+$He$_N$.}
\label{fig:massspec}
\end{figure} 
Before discussing our model calculations for these transients, let us first examine the measured time-of-flight mass spectra in more detail. Fig.~\ref{fig:massspec} (a) depicts a representative mass spectrum in the mass range around 100~amu at a pump-probe delay of 5~ps and a center wavelength $\lambda=415$ nm. The spectrum is averaged over 5000 laser shots. Clearly, the dominant fragments in this mass range are neat Rb$^+$ ions at 85 and 87 amu, where the different peak heights reflect the natural abundances of isotopes (72 and 28 \%, respectively). Even when ionizing with single laser pulses the mass spectra contain bare Rb$^+$ ions at a low level. We attribute this to a fraction of the Rb atoms desorbing off the droplets and subsequently ionizing within the laser pulse. 
A contribution to the Rb$^+$ signal may come from free Rb atoms accompanying the droplet beam as a consequence of the detachment of the Rb atom from the droplet during the pick-up process. Aside from neat Rb$^+$ atomic ions, the pump-probe mass spectra feature peaks at 89, 91, and 95 amu, which evidence the formation of Rb$^+$He and Rb$^+$He$_2$ molecular ions. These masses are usually attributed to photoionization of bound metastable Rb$^\ast$He exciplexes~\cite{Droppelmann:2004,Mudrich:2008,Fechner:2012,Loginov:2014}. 

In addition to these discrete mass peaks, we measure extended mass distributions reaching up to 64,000 amu using our time-of-flight mass spectrometer which is optimized to detecting cluster ions. These distributions are in good agreement with the size distributions of pure He nanodroplets generated in a sub-critical expansion~\cite{Lewerenz:1993,Toennies:2004}. From comparing the peak areas of the light masses Rb$^+$, Rb$^+$He$_n$, $n=1,2$ with those of the heavy droplet ions Rb$^+$He$_N$ we deduce that by ionizing with single pulses a fraction of $\lesssim 10$\% of the doped He droplets fragments into free atomic or molecular ions. The larger part of the ionized Rb-doped He droplets generates unfragmented Rb$^+$He$_N$ due to the sinking of the Rb$^+$ ion into the He droplet and the formation of a stable snowball complex~\cite{Theisen:2010}. When adding an additional time-delayed probe pulse we may expect to alter this ratio by depleting the unfragmented Rb$^+$He$_N$ fraction in favor of creating free ions Rb$^+$ and Rb$^+$He$_{1,2}$ ions after desorption.

\begin{figure}
\centering
\includegraphics[width=0.45\textwidth]{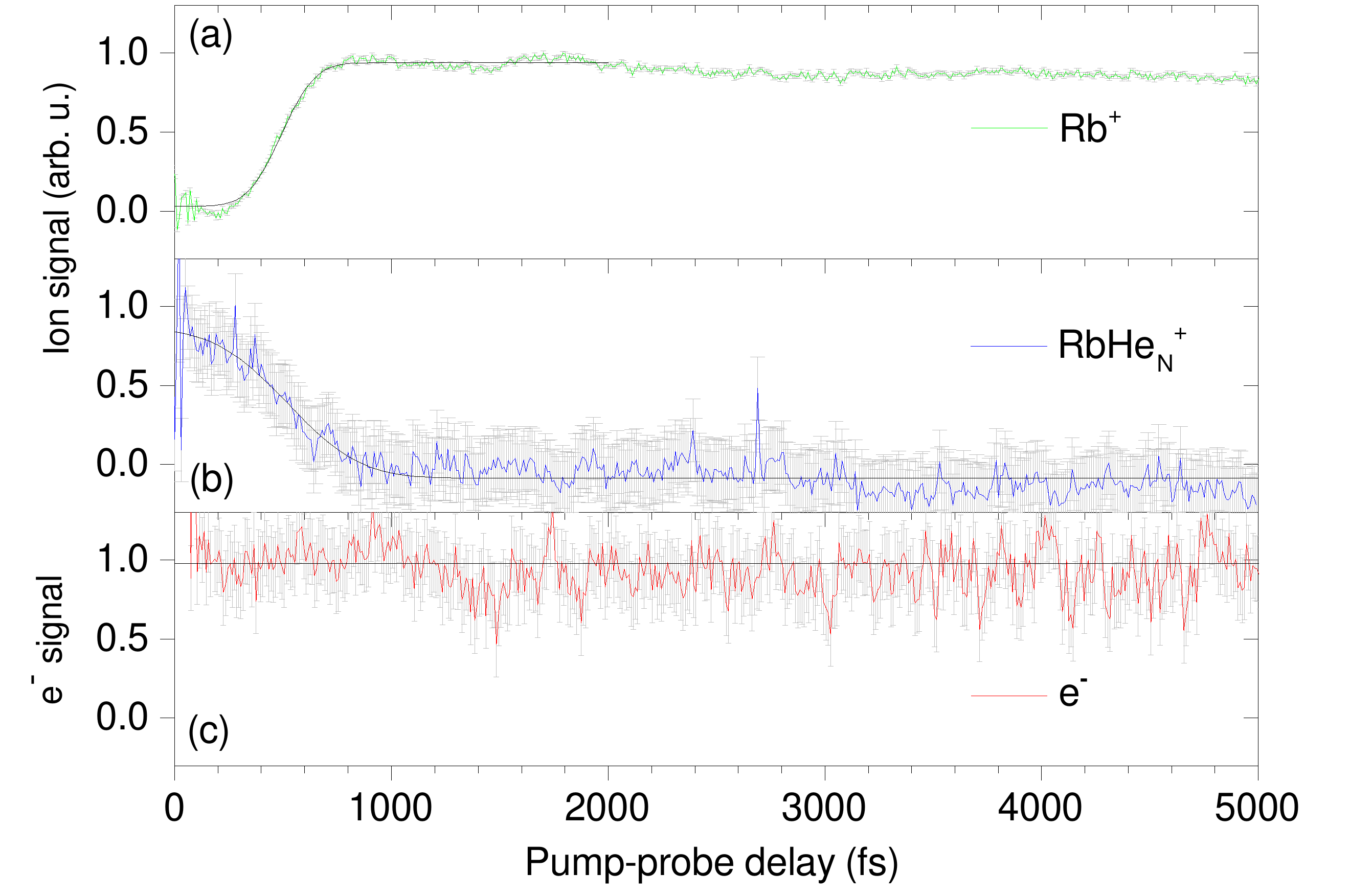}
\caption{Transient ion and electron signals measured at $\lambda=400$ nm. The ion signal traces (a and b) are obtained from integrating over the free atomic Rb$^+$ ion peaks and over the charged He droplet mass distribution, respectively. The total photoelectron signal (c) is measured using a simple electron detector. The thin smooth lines are fits to the data.}
\label{fig:triple}
\end{figure}
Indeed, the delay-dependent peak integrals of the measured mass peaks at $\lambda=400$ nm confirm this picture, see Fig.~\ref{fig:triple} (a) and (b). While the atomic Rb$^+$ ion signal sharply increases around 600 fs and remains largely constant for longer delays, the Rb$^+$He$_N$ signal displays the opposite behavior. The maximum signal level at zero delay significantly drops around $\tau =600$ fs and remains low for long delay times. 

In addition to the mass-resolved ion signals we have measured the total yield of photoelectrons, depicted in Fig.~\ref{fig:triple} (c). From comparing the electron counts with and without blocking the He droplet beam we find that for pump-probe ionization $>$79\% of photoelectrons correlates with the Rb-doped He droplet beam, $<21$\% is attributed to ionization of Rb and other species in the background gas. 
The observation that the electron count rate remains constant within the experimental scatter in the entire range of pump-probe delays indicates that the photoionization efficiency (cross-section) of a Rb atom is largely independent of its position with respect to the He droplet surface. These observations further support our interpretation of the step-like increase of Rb$^+$ counts in terms of the competition between desorption of excited Rb atoms and solvation of Rb$^+$ cations into the He droplets. 

\begin{figure}
\centering
\includegraphics[width=0.45\textwidth]{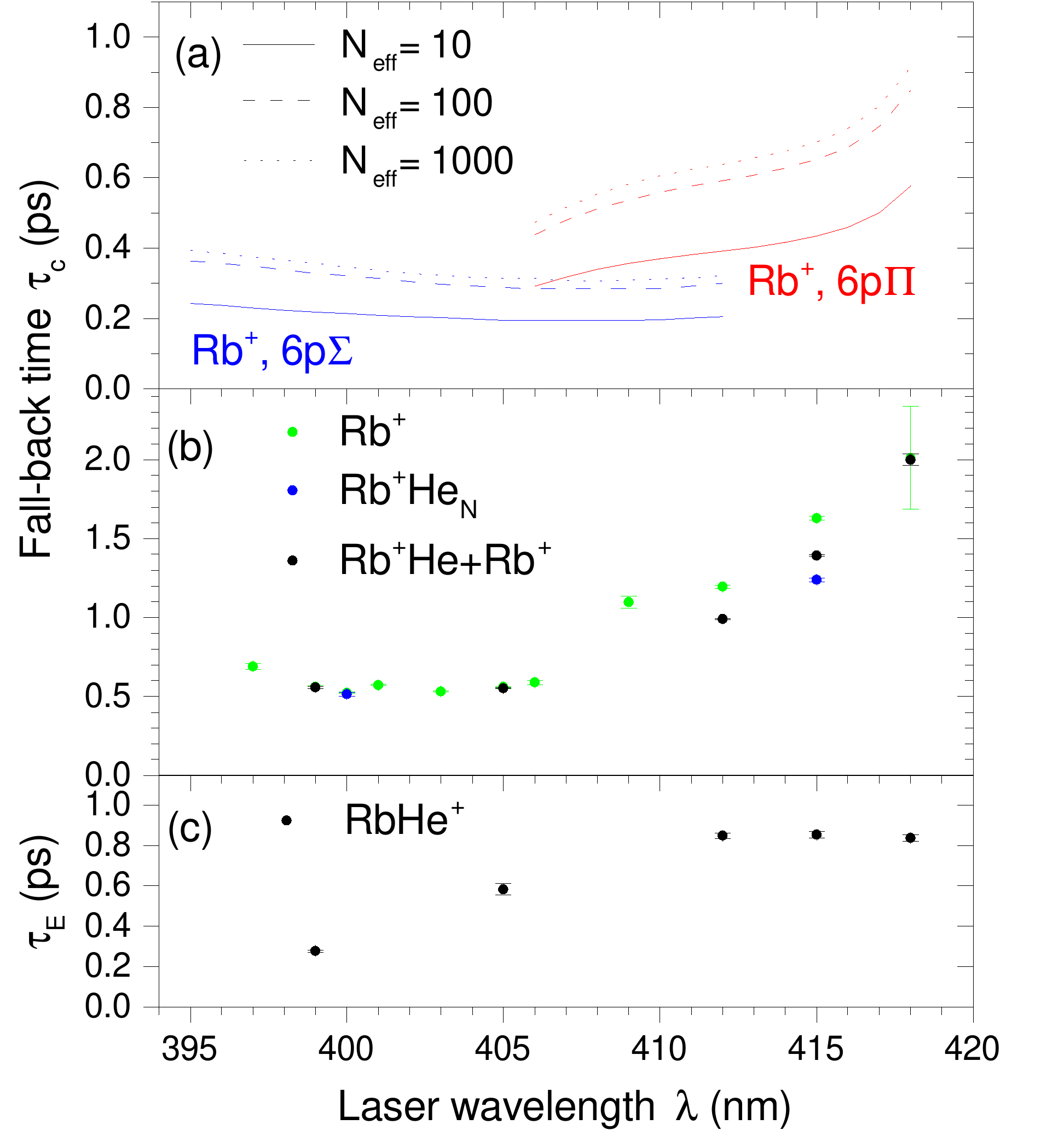}
\caption{Simulated (a) and experimental (b) fall-back times as a function of the laser wavelength, derived from the rising edges of the pump-probe transients. The curves in (a) are obtained for various effective masses $m_{\mathrm{He}_n}=N_\mathrm{eff}m_\mathrm{He}$ of the He droplet in units of the He atomic mass $m_\mathrm{He}$. The different symbols in (b) denote the experimental fit results for Rb$^+$, Rb$^+$He$+$Rb$^+$ and Rb$^+$He$_N$ signals. Panel (c) shows the exponential decay constants from fits of the Rb$^+$He ion transients with Eq.~(\ref{Fitfunction}).} 
\label{fig:tcrit}
\end{figure}

{Fig.~\ref{fig:tcrit} (b) displays a compilation of the critical delays for all measured laser wavelengths which we obtain by fitting the experimental data with an error function,
\begin{equation}
f_{Rb^+}(t)=A\cdot\{\mathrm{erf}\left[(t-\tau_c)/\sigma\right]+1\}
\label{Rb_Fitfunction}
\end{equation} 
of variable amplitude $A$, width $\sigma$ and position $\tau_c$. Shown are the results for the raw Rb$^+$ and Rb$^+$He$_N$ transients as well as those obtained by fitting the sum of the transients of Rb$^+$ atomic and Rb$^+$He molecular ions. In particular for the 6p$\Pi$ state, the latter signal more accurately reflects the dynamics of the fall-back process than the individual Rb$^+$ and Rb$^+$He transients since additional transient redistribution of population between Rb$^+$ and Rb$^+$He channels, which we discuss below, cancels out. Correspondingly, the fitted time constants of the summed Rb$^+$ and Rb$^+$He transients and those of Rb$^+$He$_N$ are in good agreement. This confirms our conception that the light ions fall back to produce heavy cluster ions at short delays. {Fig.~\ref{fig:tcrit} (c) will be discussed in section~\ref{sec:RbHeDynamics}.

\subsection{Simulations}
\label{sec:simulations}
\begin{figure}
\centering
\includegraphics[width=0.45\textwidth]{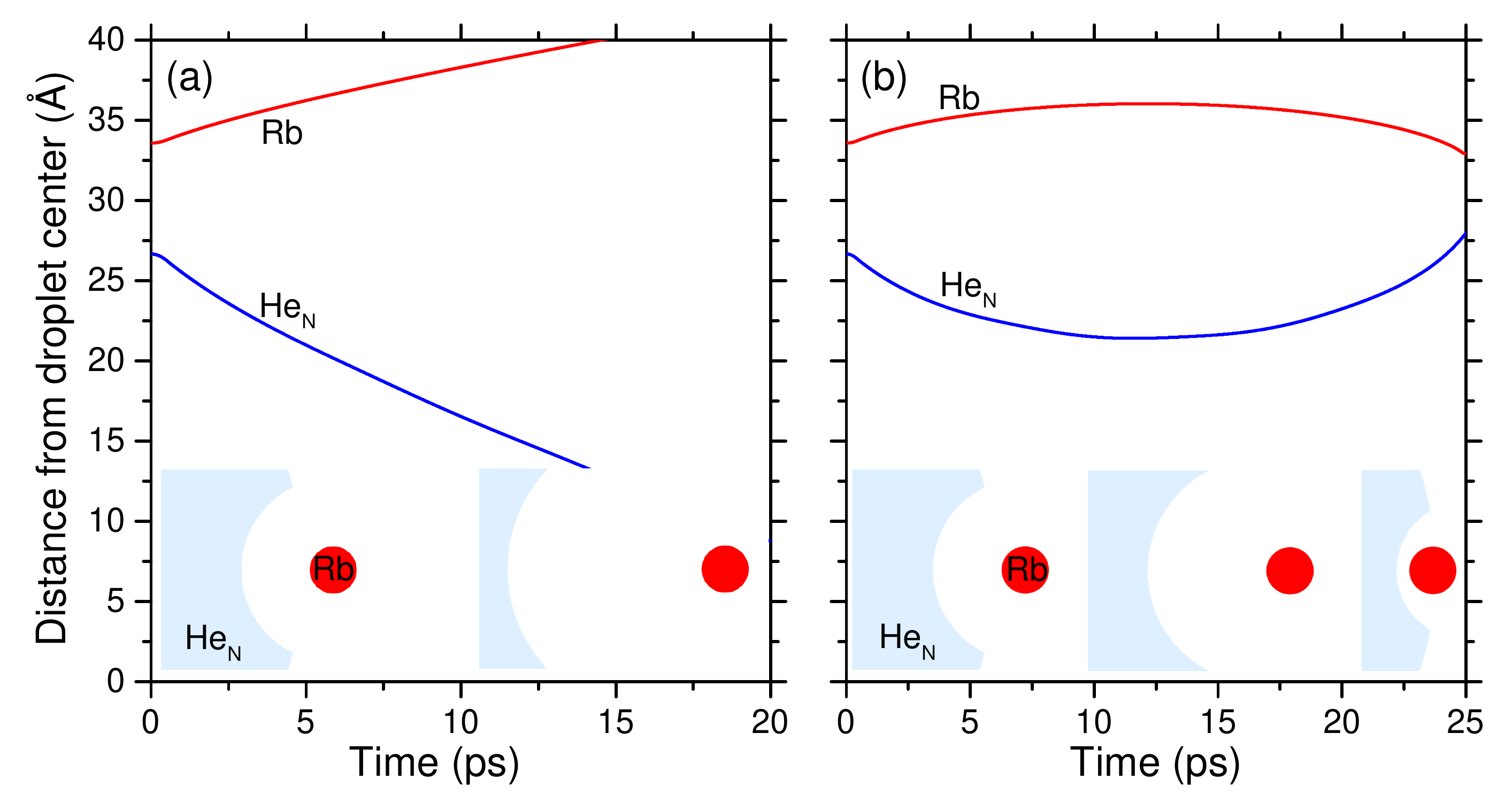}
\caption{Classical trajectories of the excited and ionized Rb atom initially located in a dimple near the He droplet surface. (a) At 6p$\Pi$-excitation ($\lambda=415$ nm) and long pump-probe delay $\tau =500$~fs the Rb atom fully desorbs off the He droplet and propagates as a free Rb$^+$ cation after ionization. (b) At shorter $\tau =400$~fs the Rb$^+$ ion turns over and falls back into the He droplet. The schemes at the bottom visualize the dynamics for increasing time from left to right.}
\label{fig:RbTrajectories}
\end{figure}



Further support for our interpretation of the experimental findings is provided by classical trajectory simulations of the dynamics of the pump-probe process. In this model, the Rb atom and the He droplet surface are taken as two point-like particles which propagate classically according to the pseudo-diatomic model potentials~\cite{Callegari:2011,Pi}. Note that these potentials were calculated based on the minimum-energy configuration of a droplet consisting of $N = 2000$ He atoms subjected to the external potential of an Rb atom in the electronic ground state.
The classical equation of motion
\begin{equation}
\mu \ddot{R} = -\frac{dV(R)}{dR},
\label{eq:Newton}
\end{equation}
is solved numerically. Here, $V=V_{\Sigma ,\,\Pi ,\,\mathrm{Rb}^+ }(R)$ denotes the potential curves of the excited and ionic states, and $R(t)$
is the distance between the Rb atom and the He dimple at the droplet surface. The initial value of the Rb-He droplet distance is the position of the minimum of the groundstate potential well (6.4~\AA).  Eq.~\ref{eq:Newton} is first solved for the neutral excited state potential $V_{\Sigma}$ or $V_\Pi$ up to the pump-probe delay time $\tau$. Subsequently, the Rb atom is considered to be ionized and the particle is propagated further using the ionic Rb$^+$-He$_N$ potential $V_{\mathrm{Rb}^+ }$. The reduced mass $\mu=m_\mathrm{Rb}m_{\mathrm{He}_n}/(m_\mathrm{Rb} + m_{\mathrm{He}_n})$ is given by the mass of the Rb atom or ion, $m_\mathrm{Rb}$, and the effective mass of the He droplet, $m_{\mathrm{He}_n}$. We set $m_{\mathrm{He}_n} = 40$~amu for the propagation of the excited as well as for the subsequent propagation of the Rb$^+$ ion with respect to the He droplet. This value is based on previous experimental as well as theoretical findings~\cite{Vangerow:2014}.

The motion of the excited and subsequently ionized Rb atom with respect to the He droplet surface is illustrated in Fig.~\ref{fig:RbTrajectories} for different initial conditions. The time-dependent positions of the Rb atom and the He surface are depicted as red and blue lines in the upper parts. The lower parts are graphical visualizations of the dynamics.
Fig.~\ref{fig:RbTrajectories} (a) depicts the case when the excitation of the Rb atom, which is initially located in the groundstate equilibrium configuration of the RbHe$_N$ complex, occurs at $t=0$ and ionization is delayed to $\tau = 500$ fs. The laser wavelength is set to $\lambda=415$ nm where the motion follows the 6p$\Pi$-potential. In this case the excited Rb atom fully desorbs off the He droplet and continues to move away from the droplet after its conversion into an ion. In the case of shorter delay $\tau = 400$~fs between excitation and ionization, shown in Fig.~\ref{fig:RbTrajectories} (b), the Rb atom turns over upon ionization as a result of Rb$^+$-He$_N$ attraction and falls back into the He droplet. 


For assessing the effect of an initial spread of Rb-He$_N$ droplet distances $R$ due to the broad laser bandwidth and of the finite length of the laser pulses $t_p$ we extend the classical trajectory calculation to a mixed quantum-classical simulation which includes an approximate description of the quantum wave packet dynamics of the system. 

The initial wave packet is obtained by transforming the spectral profile of the laser into a distribution as a function of $R$ using the potential energy difference between the initial 5s$\Sigma$ and the final 6p$\Sigma,\,\Pi$ pseudo-diatomic states. We use a Gaussian-shaped laser profile with a full width at half maximum, $\Delta\nu$, inferred from measured spectra. Typically $\Delta\nu\approx$ 2~nm, depending on the center wavelength of the laser. This corresponds to the instantaneous creation of a wave packet in the excited state centered around the minimum of the groundstate potential. 

For simulating the dynamics the wave packet is approximated by 25
segments $i$ and each segment is propagated individually according to Eq.~\ref{eq:Newton} where $R(t)$ is replaced by $R_i(t)$ representing 
the Rb-He$_N$ distance for the $i$-th segment. Convergence of the final results with respect to the number of segments has been checked.

This simplified description of the wave packet dynamics is justified because no quantum interference effects are expected for this simple dissociation reaction. Comparison with the full quantum simulation of the desorption process yields excellent agreement within the propagation time range relevant to the experiment. 

 Simulated transient yield curves as a function of the pump-probe delay $\tau$ are obtained by taking the weighted sum of the segments which have propagated outwards up to very large distances after long propagation times. This sum we identify with the fraction of desorbed atoms. Those segments which have turned over towards short distances are considered to contribute to the Rb$^+$ ions falling back into the droplet. For those segments the condition formulated initially (inequality~(\ref{eq:ineq})) is fulfilled implicitly. The finite duration of the excitation process is taken into account by convolving the resulting yield curves with the autocorrelation function of the two laser pulses.

\begin{figure}
\centering
\includegraphics[width=0.45\textwidth]{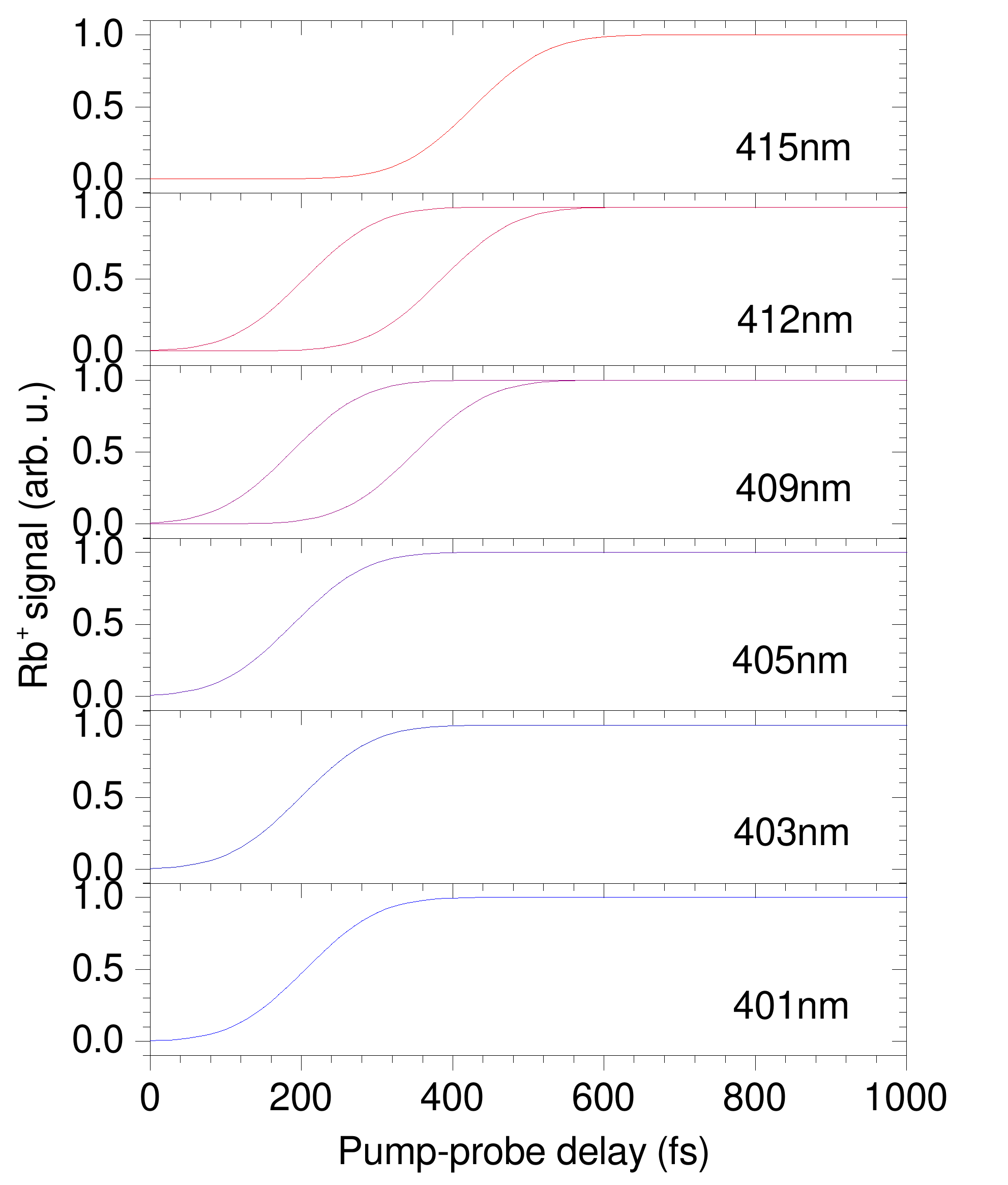}
\caption{Semiclassical simulations of the yield of free Rb$^+$ ions created by excitation and time-delayed ionization for various center wavelengths $\lambda$ of the laser pulses. See text for details.}
\label{fig:RbPPsimulation}
\end{figure}
The resulting simulated yields of free Rb$^+$ ions as a function of $\tau$ are depicted in Fig.~\ref{fig:RbPPsimulation} for various center wavelengths $\lambda$ of the laser pulses. The obtained curves qualitatively resemble the experimental ones in many respects. Excitation at long wavelengths $\lambda > 409$ nm, at which predominantly the more weakly repulsive 6p$\Pi$ state is populated, induces a smooth signal increase at about $\tau = 400$ fs. At $\lambda < 409$ nm, where predominantly the 6p$\Sigma$-state is excited, the signal rise occurs around $\tau = 210$ fs, considerably earlier than for 6p$\Pi$ excitation. This result qualitatively agrees with the experimental finding see Fig.~\ref{fig:tcrit} (b). Moreover, the superposition of the two rising edges at intermediate wavelengths $\lambda\sim 409$ nm may provide an explanation for the double-hump structure observed in the experimental Rb$^+$ transients at $\lambda < 409$ nm. However, the simulated rising edges occur at significantly shorter delay times than in the experiment, roughly by a factor 2 for excitations to the 6p$\Sigma$-state and up to a factor 4 for the 6p$\Pi$-state. 

The discrepancy between the experimental results and those of the simulations, shown in Fig.~\ref{fig:tcrit} (b), is present even when assuming very large effective masses of the He droplet $m_{\mathrm{He}_n}>1000$ amu.
We attribute this discrepancy to the limited validity of our model assumptions. In particular the interaction potentials we use were obtained on the basis of the frozen He density distribution for the RbHe$_N$ groundstate equilibrium configuration~\cite{Callegari:2011,Pi}. However, transient deformations of the He droplet surface in the course of the dynamics are likely to significantly modify the effective Rb$^\ast$-He$_N$ interactions. Recent time-dependent density functional simulations show a complex ultrafast response of the He droplet to the presence of a Rb$^+$ ion near the surface~\cite{Leal:2014}. In particular when the desorption dynamics is slow ($\Pi$-state) a complex reorganization of the He droplet surface during the Rb desorption process may be expected~\cite{Vangerow:2014}. A clear manifestation of the break-down of the simple pseudo-diatomic model is the formation of Rb$^\ast$He exciplexes which we discuss in the following section. Recently, M. Drabbels and coworkers suggested that the pseudo-diatomic potentials of the excited Na$^\ast$He$_N$ complex may be transiently shifted and even intersect~\cite{Loginov:2014,Loginov:2015}. Detailed three-dimensional simulations including the full spectrum of properties of He droplets are needed to provide an accurate description of this kind of dynamics~\cite{Hernando:2012,Mateo:2013,Vangerow:2014,Leal:2014}. Experimentally, the time evolution of the interaction potential energies will be visualized by means of fs time-resolved photoelectron spectroscopy in the near future. 

\section{R\MakeLowercase{b}H\MakeLowercase{e}$^+$ dynamics}
\label{sec:RbHeDynamics}
Aside from free Rb$^+$ ions, fs photoionization of Rb-doped He nanodroplets generates Rb$^+$He$_n$, $n=1,2$ molecular ions. Relative abundances reach up to 31\% and 1.5\%, respectively, measured at $\lambda =415$ nm corresponding to the 6p$\Pi$ excitation, see Fig.~\ref{fig:massspec}. At $\lambda =399$ nm (6p$\Sigma$-excitation), abundances are  4\% and 1\%, respectively. Free Rb$^+$He ions are associated with bound states in the Rb$^\ast$He excited states pair potentials, so called exciplexes. Both the 6p$\Sigma$ and the 6p$\Pi$-states of the RbHe diatom feature potential wells which sustain bound vibrational states that can be directly populated by laser excitation out of the groundstate of the RbHe$_N$ complex~\cite{Pascale:1983,Fechner:2012}. Thus, exciplexes are directly created in a process akin to photoassociation, in contrast to previously observed Na$^\ast$He and K$^\ast$He exciplexes which were formed by an indirect tunneling process upon excitation of the lowest p$\Pi$-states~\cite{Reho2:2000,Loginov:2015}. 


Exciplex formation is the only route to producing Rb$^+$He ions by photoionization using continuous-wave or nanosecond lasers, where ionization takes place at long delay times when the dynamics of exciplex formation and desorption off the droplets is long complete. In fs experiments, however, ionization can be triggered before or during the process of desorption of the excited atom or exciplex off the droplet surface. In this case, due to the attractive Rb$^+$-He potential a bound Rb$^+$He molecular ion can be formed upon ionization, even if the excited Rb$^\ast$-He interaction does not sustain bound states of the neutral diatom. The process of inducing a molecular bond between two initially unbound neutral species by photoionization is known as photoassociative ionization~\cite{Weiner:1990}.

\begin{figure}
\centering
\includegraphics[width=0.45\textwidth]{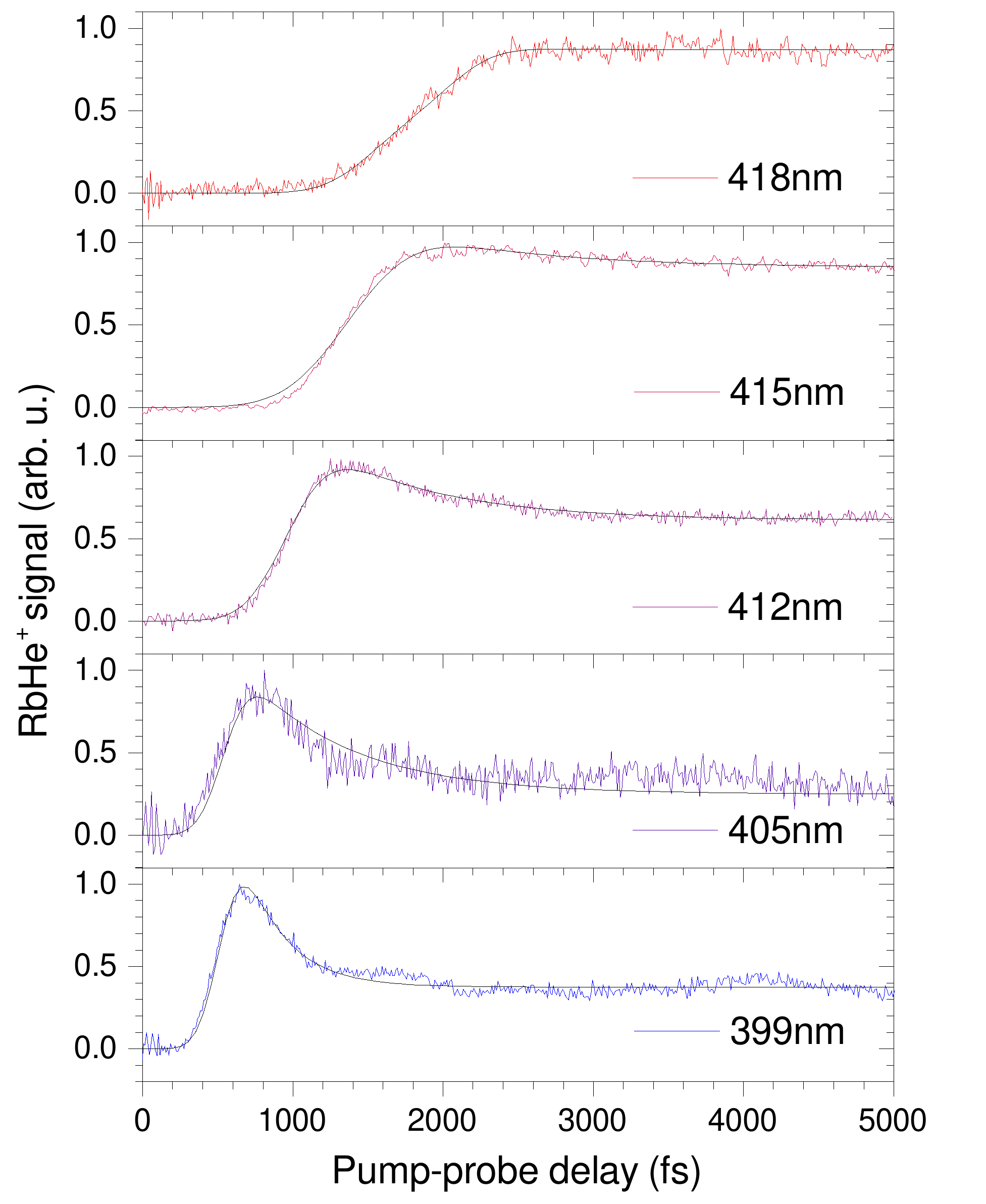}
\caption{Experimental yields of Rb$^+$He molecular ions as a function of pump-probe delay for various center wavelengths of the laser pulses. The thin smooth lines are fits to the data.} 
\label{fig:RbHeTransients}
\end{figure}
\subsection{Experimental results}
The transient yield of Rb$^+$He for various laser wavelengths is displayed in Fig.~\ref{fig:RbHeTransients}. Similarly to the Rb$^+$ transients, we measure vanishing Rb$^+$He pump-probe signal contrast around zero delay. For increasing laser wavelength from $\lambda =399$ up to 418~nm, which corresponds to the crossover from the 6p$\Sigma$ to the 6p$\Pi$ excited pseudo-diatomic states, a step-like increase of the Rb$^+$He ion signal occurs at delays ranging from $\tau =500$~fs up to about $2000$~fs. Besides, at $\lambda\lesssim 415$ nm we measure a transient overshoot of the Rb$^+$He signal by up to about 100\% of the signal level at long delays. The transient yield of Rb$^+$He is fitted using the model function
\begin{equation}
f_{Rb^+He}(t)=f_{Rb^+}(t)(Ee^{-t/\tau_E}+1).
\label{Fitfunction}
\end{equation}  
As for the Rb$^+$ case, $f_{Rb^+}(t)$ models the fall-back dynamics by Eq.~\ref{Rb_Fitfunction}. Additionally, the exponential function with amplitude $E$ and time constant $\tau_E$ takes the transient overshoot into account, whereas the additive constant account for a $\tau$-independent Rb$^+$He formation channel. The exponential time constants $\tau_E$ are plotted as black circles in Fig.~\ref{fig:tcrit} (c). To obtain these values, the parameters $\tau_c$ and $\sigma$ are taken as constants from the fit of the sum of Rb$^+$ and Rb$^+$He signals with Eq.~\ref{Rb_Fitfunction}. Here we make the assumption that the fall-back dynamics is only weakly perturbed by the attachment of a He atom to the Rb atom or ion, which is confirmed by our simulations.

\begin{figure}
\centering
\includegraphics[width=0.45\textwidth]{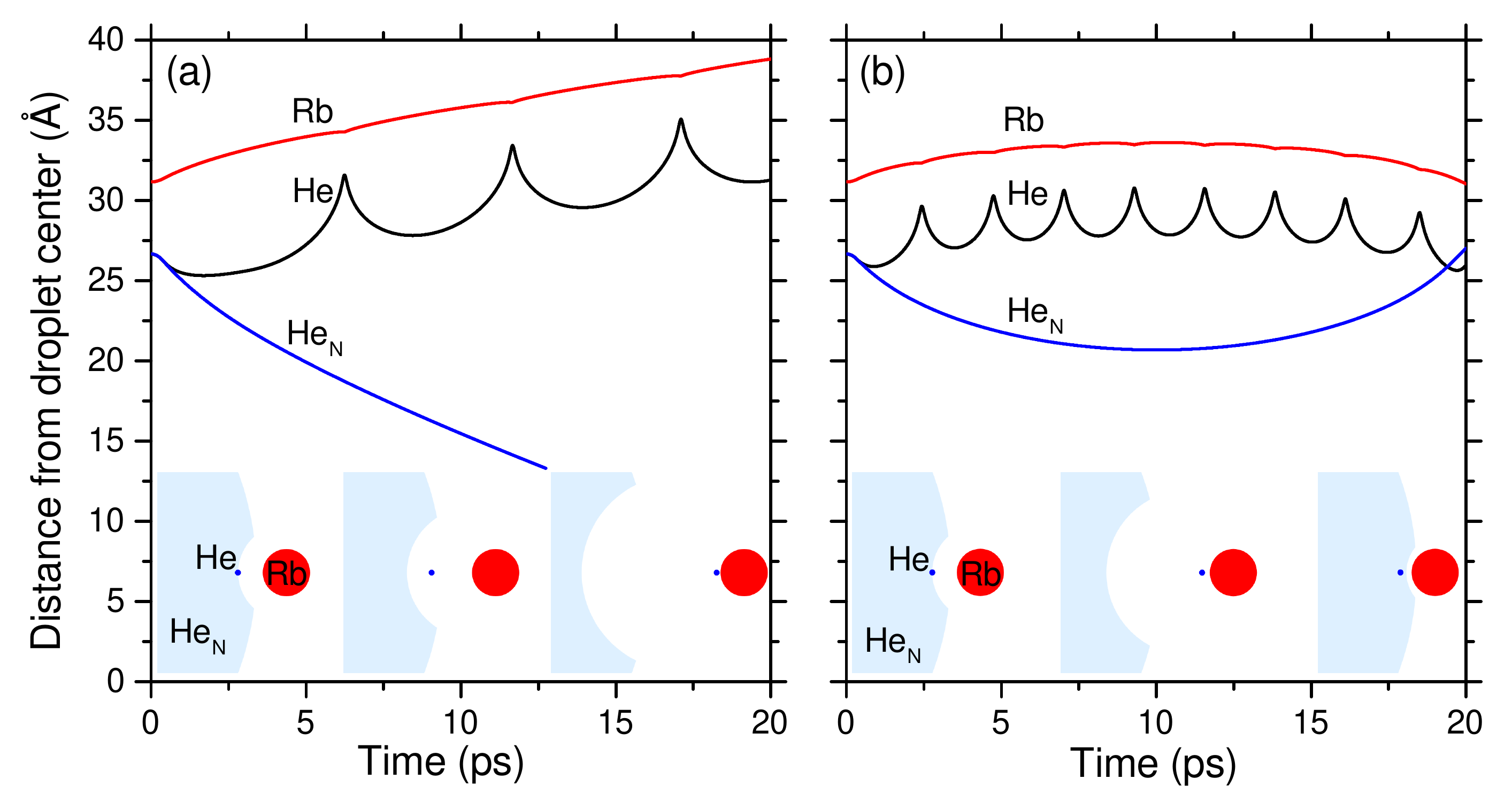}
\caption{Classical trajectories of the Rb-He-He$_N$ three-body system at $\lambda=409$ nm. The schemes at the bottom visualize the various dynamics for increasing time from left to right. (a) The excited Rb atom departing from the He droplet surface suddenly experiences Rb$^+$He pair attraction upon ionization at $\tau =400$~fs (a). Consequently, a He atom attaches to the Rb atom while it leaves the droplet. (b) For short delay $\tau=350$~fs at $\lambda=409$~nm a Rb$^+$He molecule forms as in (a) but the attraction towards the He droplet makes it turn over and fall back.}
\label{fig:RbHeTrajectories}
\end{figure}

\subsection{Simulation of photoassociative ionization}
For a more quantitative interpretation of the Rb$^+$He transients we extend our classical and mixed quantum-classical models to the one-dimensional three-body problem in the second stage of the calculation  after ionization has occurred by including one individual He atom out of the surface layer. The
classical trajectories are now obtained by solving three individual coupled equations of motion for the three individual particles Rb$^+$, He, and He$_n$. The Rb$^\ast$-He$_N$ interaction leading to desorption is represented by the pseudodiatomic potentials as before. The Rb$^+$-He dynamics is
initialized by the velocity and distance of the dissociating Rb$^\ast$He$_N$ complex at the moment of ionization. The Rb$^+$-He pair interaction is given by the Rb$^+$-He pair potential~\cite{Koutselos:1990} augmented by a 16.7~cm$^{-1}$ deep potential step to account for the He-He$_N$ extraction energy as suggested by Reho et al.~\cite{Reho2:2000,Droppelmann:2004,Fechner:2012}.\\  


Exemplary trajectories are shown in Fig.~\ref{fig:RbHeTrajectories} for two cases at $\lambda=409$ nm. For long pump-probe delays the Rb$^+$ ion leaves the He droplet without attaching a He-atom, as shown in Fig.~\ref{fig:RbTrajectories}. However, there is a range of delays in which the desorbing Rb atom is far enough away from the droplet so that it will not fall back upon ionization, but it is still close enough to attract a He atom out of the droplet surface so as to form a bound molecular ion by PAI (Fig.~\ref{fig:RbHeTrajectories} (a)). Fig.~\ref{fig:RbHeTrajectories} (b) illustrates the dynamics at short delay when the attractive forces acting between the Rb$^+$ ion and the droplet surface prevent the full desorption and Rb$^+$-He pairwise attraction leads to the formation of Rb$^+$He.\\


\begin{figure}
\centering
\includegraphics[width=0.45\textwidth]{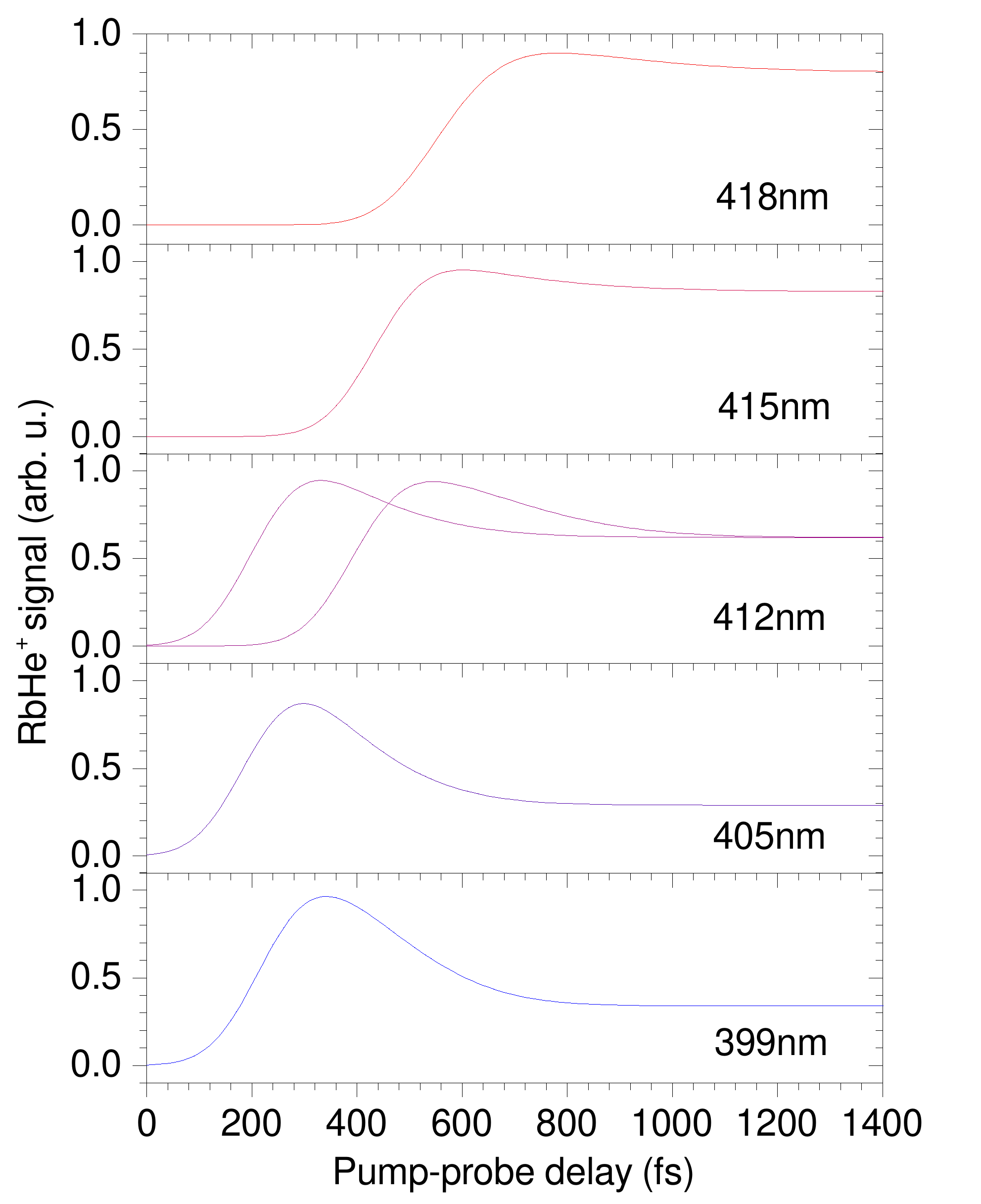}
\caption{Simulations of the yield of free Rb$^+$He-molecules created by excitation and time-delayed ionization for various center wavelengths of the laser pulses. See text for details.}
\label{fig:RbHePPsimulation}
\end{figure}

For simulating the transient Rb$^+$He yields to compare with the experimental data shown in Fig.~\ref{fig:RbHeTransients} we extend the mixed quantum-classical model for the desorption dynamics of bare Rb described in Sec.~\ref{sec:simulations}. It is augmented by computing the probability of populating bound vibrational states of the Rb$^+$He molecule for each segment of the Rb wave packet upon ionization as the sum of spatial overlap integrals
\begin{align}
	p_i^{PAI}(\tau )=\sum_v\left|\int^\infty_\infty
	\phi_v (R) \cdot \psi_i(R,\tau ) \; dR\right|^2.
\end{align}
Here $\psi^i$ denotes the $i$-th wave packet segment and $\phi_v$ stands for the vibrational wave functions of Rb$^+$He calculated using R. J. LeRoy's LEVEL program~\cite{level} for the 6p$\Sigma ,\,\Pi$ pair potentials of Rb$^+$He~\cite{Pascale:1983}. The identification of bound and free Rb$^+$He ions in the simulation is based on analyzing the final Rb$^+$-He and Rb$^+$He-He$_N$ distances after long delays $\tau >10$~ps, respectively. The final probability $P$ of detecting a Rb$^+$He molecule is obtained by summing up the detection probabilities for every segment,
\begin{equation}
	P(\tau ) = \sum_i p_i^D(\tau)\cdot (p_i^{PAI}(\tau ) + p_{ex}).
\label{eq:Ptau}
\end{equation}
In agreement with Eq. \ref{Fitfunction}, $p_i^D$ denotes the desorption probability and $p_{ex}$ is the probability of creating a bound neutral Rb$^\ast$He exciplex, which is assumed to occur instantaneously upon laser excitation and thus does not depend on $\tau$. Since the relative contributions of PAI and direct exciplex formation are not precisely known we resign from quantitatively modeling the relative efficiencies of the two pathways leading to free Rb$^+$He. Instead we adjust them to the experimental transients by taking $p_{ex}$ as a free fit parameter. The transient signal $P(\tau )$ is finally convoluted with the intensity autocorrelation function of the laser pulses, as for the Rb$^+$ transients. 

The resulting simulated yields of free Rb$^+$He molecular ions are depicted in Fig.~\ref{fig:RbHePPsimulation}. Clearly, the same general trends as for neat Rb$^+$ ions are recovered: (i) at short delay times $\tau < 200$ fs the appearance of Rb$^+$He is suppressed due to the falling back of the ion into the He droplet; (ii) longer laser wavelengths $\lambda\gtrsim 409$ nm (6p$\Pi$-excitation) lead to weaker repulsion and therefore to the delayed appearance of free ions as compared to 6p$\Sigma$-excitation at $\lambda\lesssim 409$ nm. These results again qualitatively agree with the experimental findings but the simulated appearance times are shorter by a factor 2-4, as shown in Fig.~\ref{fig:tcrit}. As in the Rb$^+$ case we attribute these deviations to the use of pseudo-diatomic potentials calculated for the frozen RbHe$_N$ groundstate complex.

Moreover, the simulation reproduces a signal overshoot around $\tau=300$ fs at short wavelengths, which is due to the contribution of the photoassociative ionization channel. Association of a bound Rb$^+$He ion is possible only at sufficiently short Rb-He distances given at delays $\tau\lesssim 600$ fs for 6p$\Sigma$-excitation and $\tau\lesssim 900$ fs for 6p$\Pi$-excitation, respectively. At these delay times, the PAI signal adds to the signal due to ionization of Rb$^\ast$He exciplexes formed directly by the pump pulse. Note that we have adjusted to the experimental curves the relative contributions to the total Rb$^+$He signal arising from PAI and from exciplex ionization. Therefore our model does not permit a quantitative comparison with the experimentally measured signal amplitudes. A more detailed three-dimensional simulation of the dynamics is needed for a fully quantitative interpretation~\cite{Mateo:2013}. Nevertheless, we take the simulation result as a clear indication that PAI is an additional channel producing He-containing ionic complexes which needs to be considered in experiments involving photoionization of dopants attached to He droplets when using ultrashort pulses. 


We note that in the particular case of exciting into the 6p$\Sigma$-state of the RbHe$_N$ complex, unusual Rb$^+$He signal transients may arise from the peculiar shape of the RbHe pair potential curve which features a local potential maximum at intermediate Rb-He distance~\cite{Pascale:1983,Fechner:2012}. This potential barrier causes the highest-lying Rb$^\ast$He vibrational states to be predissociative. Semi-classical estimates yield predissociation time constants for the two highest vibrational levels $v=5$ and $6$ of 3.5 ns and 2.2 ps, respectively. However, these values significantly exceed the exponential decay times inferred from the measured transients (see Fig.~\ref{fig:tcrit} (c)). Moreover, we may expect that not only the highest vibrational levels are populated. Note that for the case of Na$^\ast$He and K$^\ast$He formed in the lowest 3p$\Pi$ and 4p$\Pi$ states, respectively, all vibrational levels including the lowest ones were found to be populated to varying extents depending on the laser wavelength~\cite{Reho:2000}. Therefore we tend to discard predissociation of Rb$^\ast$He exciplexes as the origin of the peculiar shape of the Rb$^+$He transients, although we cannot strictly rule it out. More insight into the Rb$^+$He dynamics may be provided by further measurements using electron and ion imaging detection.

\section{Summary}
This experimental fs pump-probe study discloses the competing dynamics of desolvation and solvation of excited and ionized states, respectively, of Rb atoms which are initially located at the surface of He nanodroplets. The generic feature of the pump-probe transients -- the time-delayed appearance of photoions -- is shown to result from the falling back of ions into the droplets when the ionization occurs at an early stage of the desorption process. This interpretation is backed by the experimental observation of the opposing trend when measuring the yield of unfragmented He nanodroplets containing a Rb$^+$ ion. Furthermore, mixed quantum-classical model calculations based on one-dimensional pseudo-diatomic potentials confirm this picture qualitatively. The limited quantitative agreement with the experimental results is attributed to the use of model potentials in the calculations, which do not account for the transient response of the He density upon excitation and ionization of the Rb dopant atom. Much better agreement may be expected from three-dimensional time-dependent density functional simulations~\cite{Vangerow:2014,Leal:2014} of the full pump-probe sequence which are currently in preparation.

Pump-probe dynamics similar to the Rb$^+$ case is observed when detecting Rb$^+$He molecular ions which primarily result from photoionization of Rb$^\ast$He exciplexes. The peculiar structure of the Rb$^+$He transients as well as extended model calculations indicate that photoassociative ionization is an additional mechanism of forming He-containing ionic complexes in fs experiments. However, the dynamics resulting from the additional photoassociative ionization channel cannot unambiguously be distinguished from predissociation of Rb$^\ast$He exciplexes in high-lying vibrational levels of the 6p$\Sigma$-state. 

These results shed new light on the interpretation of the Rb$^+$He pump-probe transients measured previously by ionizing via the lowest 5p$\Pi$ excited state~\cite{Droppelmann:2004,Mudrich:2008}. The signal increase at short delays was interpreted as the manifestation of the formation dynamics of the Rb$^\ast$He exciplex by a tunnelling process. Possibly the competing desorption of the excited neutral and the fall-back of the photoion actually more crucially determines the rise time of the Rb$^+$He signal in those transients. This issue will be elucidated in future experiments using two-color pump-probe ionization via the 5p droplet states and at low laser repetition rate so as to exclude concurrent effects by subsequent laser pulses. 
Furthermore, we will investigate the photodynamics of metal atom-doped He nanodroplets in more detail by applying refined detection schemes such as ion and electron imaging~\cite{Fechner:2012,Vangerow:2014} and coincidence detection~\cite{Buchta:2013,BuchtaJCP:2013}. 

\begin{acknowledgments}
The authors gratefully acknowledge fruitful discussions with W. Strunz. We thank M. Pi and M. Barranco for providing us with the Rb$^+$-He$_{2000}$ pseudodiatomic potential curve. Furthermore, we thank the Deutsche Forschungsgemeinschaft for financial support.
\end{acknowledgments}



%


\end{document}